\documentstyle[psfig,graphicx]{mn}
\newif\referee
\newif\ifAMStwofonts

\catcode`\@=11
\def\gsim{\ifmmode{\mathrel{\mathpalette\@versim>}}
    \else{$\mathrel{\mathpalette\@versim>}$}\fi}
\def\lsim{\ifmmode{\mathrel{\mathpalette\@versim<}}
    \else{$\mathrel{\mathpalette\@versim<}$}\fi}
\def\@versim#1#2{\lower 2.9truept \vbox{\baselineskip 0pt \lineskip
    0.5truept \ialign{$\m@th#1\hfil##\hfil$\crcr#2\crcr\sim\crcr}}}
\catcode`\@=12

\def\ml{\Upsilon_*}
\def\Mstar{M_*}
\def\Psistar{\Psi_*}
\def\rhostar{\rho_*}
\def\rhoM{\rho_{\rm M}}
\def\Lb{L_{\rm B}}
\def\Lbsol{L_{\rm B\odot}}
\def\Msol{M_{\odot}}
\def\Mbh{M_{\rm BH}}
\def\Mbhzero{M_{\rm BH,0}}
\def\Ie{\langle I\rangle _{\rm e}}
\def\Re{R_{\rm e}}
\def\cRe{\langle R\rangle _{\rm e}}
\def\Dt{\Delta t}
\def\Dtmax{\Delta t_{\rm max}}
\def\Dtmin{\Delta t_{\rm min}}
\def\sae{a_{\rm e}}
\def\sbe{b_{\rm e}}
\def\rc{r_{\rm c}}

\def\rv{r_{\rm V}} 
\def\rvzero{r_{\rm V,0}} 
\def\rbh{r_{\rm BH}}
\def\rme{r_{\rm M}}
\def\r90{\langle r\rangle _{90}}
\def\runo{\langle r\rangle _{90,1}}
\def\rdue{\langle r\rangle _{90,2}}

\def\sg0{\sigma_0}
\def\sgv{\sigma_{\rm V}}
\def\sgvzero{\sigma_{\rm V,0}}
\def\gm{\gamma}
\def\Td{T_{\rm dyn}}

\def\en{{\mathcal{E}}}
\def\psit{\Psi_{\rm T}}
\def\ku{k_1}
\def\kd{k_2}
\def\kt{k_3}

\def\alphatol{\alpha_{\rm tol}}
\def\thetamin{\theta_{\rm min}}
\ifoldfss
  \ifCUPmtlplainloaded \else
    \NewTextAlphabet{textbfit} {cmbxti10} {}
    \NewTextAlphabet{textbfss} {cmssbx10} {}
    \NewMathAlphabet{mathbfit} {cmbxti10} {} 
    \NewMathAlphabet{mathbfss} {cmssbx10} {} 
  \fi
  \ifAMStwofonts
    \ifCUPmtlplainloaded \else
      \NewSymbolFont{upmath} {eurm10}
      \NewSymbolFont{AMSa} {msam10}
      \NewMathSymbol{\upi}     {0}{upmath}{19}
      \NewMathSymbol{\umu}     {0}{upmath}{16}
      \NewMathSymbol{\upaellitrtial}{0}{upmath}{40}
      \NewMathSymbol{\leqslant}{3}{AMSa}{36}
      \NewMathSymbol{\geqslant}{3}{AMSa}{3E}

      \let\leq=\leqslant 
      \let\geq=\geqslant 
    \fi
  \fi
\fi 

\ifnfssone
  \newmathalphabet{\mathit}
  \addtoversion{normal}{\mathit}{cmr}{m}{it}
  \addtoversion{bold}{\mathit}{cmr}{bx}{it}
  \newmathalphabet{\mathbfit} 
  \addtoversion{normal}{\mathbfit}{cmr}{bx}{it}
  \addtoversion{bold}{\mathbfit}{cmr}{bx}{it}
  \newmathalphabet{\mathbfss} 
  \addtoversion{normal}{\mathbfss}{cmss}{bx}{n}
  \addtoversion{bold}{\mathbfss}{cmss}{bx}{n}
  \ifAMStwofonts
    \ifCUPmtlplainloaded \else
      \UseAMStwoboldmath
      \makeatletter
      \new@mathgroup\upmath@group
      \define@mathgroup\mv@normal\upmath@group{eur}{m}{n}
      \define@mathgroup\mv@bold\upmath@group{eur}{b}{n}
      \edef\UPM{\hexnumber\upmath@group}
      \new@mathgroup\amsa@group
      \define@mathgroup\mv@normal\amsa@group{msa}{m}{n}
      \define@mathgroup\mv@bold\amsa@group{msa}{m}{n}
      \edef\AMSa{\hexnumber\amsa@group}
      \makeatother
      \mathchardef\upi="0\UPM19
      \mathchardef\umu="0\UPM16
      \mathchardef\upartial="0\UPM40
      \mathchardef\leqslant="3\AMSa36
      \mathchardef\geqslant="3\AMSa3E

      \let\leq=\leqslant 
      \let\geq=\geqslant 
    \fi
  \fi
\fi 

\ifnfsstwo
  \DeclareMathAlphabet{\mathbfit}{OT1}{cmr}{bx}{it}
  \SetMathAlphabet\mathbfit{bold}{OT1}{cmr}{bx}{it}
  \DeclareMathAlphabet{\mathbfss}{OT1}{cmss}{bx}{n}
  \SetMathAlphabet\mathbfss{bold}{OT1}{cmss}{bx}{n}
  \ifAMStwofonts
    \ifCUPmtlplainloaded \else
      \DeclareSymbolFont{UPM}{U}{eur}{m}{n}
      \SetSymbolFont{UPM}{bold}{U}{eur}{b}{n}
      \DeclareSymbolFont{AMSa}{U}{msa}{m}{n}
      \DeclareMathSymbol{\upi}{0}{UPM}{"19}
      \DeclareMathSymbol{\umu}{0}{UPM}{"16}
      \DeclareMathSymbol{\upartial}{0}{UPM}{"40}
      \DeclareMathSymbol{\leqslant}{3}{AMSa}{"36}
      \DeclareMathSymbol{\geqslant}{3}{AMSa}{"3E}

      \let\leq=\leqslant 
      \let\geq=\geqslant 
    \fi
  \fi
\fi 

\ifCUPmtlplainloaded \else
  \ifAMStwofonts \else 
    \def\upi{\pi}
    \def\umu{\mu}
    \def\upartial{\partial}
  \fi
\fi

\title{Galaxy merging, the Fundamental
       Plane of elliptical galaxies, and the $\Mbh$-$\sg0$ relation}
\author[C. Nipoti, P. Londrillo and L. Ciotti]
  {C.~Nipoti,$^1$ P.~Londrillo$^2$ and L.~Ciotti$^{1,3}$\\
  $^1$Dipartimento di Astronomia, Universit\`a di Bologna, 
      via Ranzani 1, 40127 Bologna, Italy\\
  $^2$INAF - Osservatorio Astronomico di Bologna, 
      via Ranzani 1, 40127 Bologna, Italy\\
  $^3$Scuola Normale Superiore, 
      Piazza dei Cavalieri 7, 56126 Pisa, Italy}
  
\date{Accepted version 19/02/2003, submitted in original form 16/09/2002}

\pagerange{\pageref{firstpage}--\pageref{lastpage}}
\pubyear{2003}

\begin{document}

\maketitle

\label{firstpage}

\begin{abstract}

We explore the effects of {\it dissipationless} merging on the
Fundamental Plane of elliptical galaxies by using a N-body code based
on a new, high performance numerical scheme (Dehnen 2002).  We
investigate the two extreme cases of galaxy growth by {\it equal mass
merging} and {\it accretion} of small stellar systems; in a subset of
simulations we also consider the presence of dark matter halos around
the merging galaxies. Curiously, we found that the Fundamental Plane
is preserved by major merging, while in the accretion scenario its
edge--on thickness is only marginally reproduced, with substantial
thickening in the case of merging with low angular momentum. We also
found that both the Faber-Jackson and Kormendy relations are {\it not}
reproduced by the simulations, in accordance with the results of a
preliminary analysis based on a simple application of the virial
theorem.  Finally, we discuss the implications of our results for the
origin of the $\Mbh$-$\sg0$ and Magorrian relations. We found that
dissipationless merging is unable to reproduce the $\Mbh$-$\sg0$
relation, if the black hole masses add linearly (while the Magorrian
relation is nicely reproduced); on the contrary a black hole merging
with substantial emission of gravitational waves reproduces the
$\Mbh$-$\sg0$ relation but fails at reproducing the Magorrian
relation. We argue that our results strongly point towards a major
role of {\it dissipation} in the formation of early--type galaxies and
in the growth of their central supermassive black holes, thus
supporting the idea of a link between galaxy formation and QSO
activity.

\end{abstract}


\section{Introduction}

Two distinct scenarios have been proposed to describe the formation of
elliptical galaxies (Es): roughly speaking, according to the {\it
monolithic scenario} Es form at high redshift in dissipative 
collapses (see, e.g., Eggen, Lynden-Bell \& Sandage 1962, Larson
1975), while in the {\it hierarchical merging scenario} spheroidal
systems are the end--products of several merging processes of smaller
galaxies, the last major merger taking place in relatively recent
times, i.e. at $z \lsim 1$ (see, e.g., White \& Rees 1978, Kauffmann
1996, Cole et al. 2000). In particular, this second picture is
supported by some observational data suggesting that a fraction of red
galaxies in clusters at intermediate redshift are undergoing merging
processes: these galaxies could be the progenitors of present--day
early--type galaxies (van Dokkum et al. 1999).

In any case, it is well known that Es satisfy many scaling relations
(more or less tight) that must be accounted for by any proposed
formation scenario: for example the Faber--Jackson relation (hereafter
FJ, Faber \& Jackson 1976), the Kormendy relation (Kormendy 1977), the
${\rm Mg}_2$-$\sg0$ relation (see, e.g., Burstein et al. 1988, Bender,
Burstein \& Faber 1993) the color--magnitude relation (Bower, Lucey \&
Ellis 1992), and the very tight Fundamental Plane (hereafter FP;
Djorgovski \& Davis 1987, Dressler et al. 1987).  In addition, it is
now widely accepted that central supermassive black holes (hereafter
BHs) are a common characteristic of spheroidal stellar systems (see,
e.g., Kormendy \& Richstone 1995, van der Marel 1999, de Zeeuw 2001);
it has also been found that their mass ($\Mbh$) is linearly
proportional to the mass of the host spheroid (Magorrian et al. 1998),
and it is even more correlated with the spheroid central stellar
velocity dispersion (Gebhardt et al.  2000, Ferrarese \& Merritt
2000). Due to their relevance in this work, here we briefly review the
main properties of the FP, FJ, Kormendy and $\Mbh$-$\sg0$ relations.

The FP of Es is a scaling relation between three of their basic {\it
observational} properties, namely the circularized effective radius
{$\cRe\equiv\sqrt{\sae\sbe}$} (where {$\sae$} and {$\sbe$} are the
major and minor semi-axis of the effective isophotal ellipse), the
central velocity dispersion {$\sg0$}, and the mean effective surface
brightness {$\Ie{\equiv}\Lb/2\pi{\cRe}^2$} (where {$\Lb$} is the
luminosity of the galaxy, for example in the Johnson B-band).  A
parameterization of the FP, that we adopt in this paper, has been
introduced by Bender, Burstein \& Faber (1992, hereafter BBF), by
defining the three variables
\begin{equation}
\ku\equiv\frac{\log\sg0^2 + \log\cRe}{\sqrt 2},
\end{equation}
\begin{equation}
\kd\equiv\frac{\log\sg0^2 + 2\log\Ie - \log\cRe}{\sqrt 6},
\end{equation}
\begin{equation}
\kt\equiv\frac{\log\sg0^2 - \log\Ie -\log\cRe}{\sqrt 3}.
\end{equation}
In particular, when projected on the $(\ku ,\kt)$ plane, the FP is
seen almost edge--on and it is considerably thin, while the
distribution of galaxies in the $(\ku ,\kd)$ plane is
broader.  For example, Virgo ellipticals studied by BBF are
distributed on the $(\ku ,\kt)$ plane according to the best--fit
relation
\begin{equation}
\kt=0.15\ku+0.36,
\end{equation}
(when adopting respectively, kpc, km s$^{-1}$ and $\Lbsol$ pc$^{-2}$
as length, velocity and surface brightness units), with a very small
1-$\sigma$ dispersion of ${\rm rms}(\kt)\simeq 0.05$ over the whole
range spanned by the data, $2.6\, \lsim\, \ku\, \lsim\, 4.6$ ($0.75\,
\lsim \, \kt\, \lsim \, 1.05$). This translates in a scatter of
$\simeq15\%$ in $\cRe$ for fixed $\Lb$ and $\sg0$. The slope and the
scatter of equation (4) are usually called ``tilt'' and ``thickness''
of the FP, respectively. By combining equations (1) and (3) with
equation (4) the FP relation is obtained in terms of the observables,
and is in good agreement with the FP derived (in the same photometric
band and for a much larger galaxy sample) by J{\o}rgensen, Franx \&
Kj{\ae}rgaard (1996)\footnote{As it is well known, the FP tilt
depends on the adopted photometric band. Curiously, when using the
$k$-space, the coefficient in the K-band is only slightly different
from 0.15, with a reported value of 0.147 (Pahre, Djorgovski \& de
Carvalho 1998b).}.

Strictly related to the FP are the less tight FJ and Kormendy
relations. These scaling relations have been often interpreted just as
projections of the FP with no additional information, and, for this
reason, their constraining power on galaxy formation scenarios has
been underestimated. However this is misleading, because these two
relations describe {\it how} Es are distributed on the FP and so, even
if characterized by a larger scatter than the edge--on view of the FP,
they contain important information on the galactic properties.  The
proposed original form of the FJ relation was $\Lb \propto \sg0^n$,
with $n \approx 4$, while Davies et al. (1983) found that the
double--slope fit
\begin{equation}
{\Lb \over 10^{11} \Lbsol} \simeq 0.23\left( {\sg0 \over 300 \, {\rm
km\,s^{-1}}}\right)^{2.4}+0.62\left( {\sg0 \over 300 \, {\rm
km\,s^{-1}}}\right)^{4.2}
\end{equation}
provides a better description of the data.  Note that for small
galaxies and bulges ($\sg0 \, \lsim \, 170 \,{\rm km\,s^{-1}}$) the
exponent is $\sim 2.4$, considerably smaller than 4; this situation is
also reflected by the fit of Dressler et al. (1987) to the galaxies of
Virgo cluster, $\Lb\propto\sg0^{3.5}$. Due to indeterminacy of the
exact value of $n$, in the rest of the paper we use 3.5 and 4 as
representative values of the FJ exponent.

The total luminosity of spheroidal systems also correlates with their
length scale as measured by $\cRe$: in fact, the Kormendy relation can
be written in the form
\begin{equation}
\cRe \propto \Lb^{a},
\end{equation}
where the exponent $a$ is strongly dependent on the galaxy sample
used, and is found in the range $0.88 \, \lsim \, a \, \lsim \, 1.62$
(Ziegler et al. 1999). The latest estimates seem to converge to a
value $a\sim0.7$ or less, as a function of waveband (Bernardi et
al. 2003).

In the last years, another important scaling relation has been added
to the list, the so--called $\Mbh$-$\sg0$ relation (see, e.g.,
Gebhardt et al.  2000; Ferrarese \& Merritt 2000, Merritt \& Ferrarese
2001, Tremaine et al. 2002):
\begin{equation}
\Mbh \propto \sg0^{\alpha},
\end{equation}
where $\sg0$ is the projected central velocity dispersion of the
parent galaxies (or within $\cRe$), and the exact value of $\alpha$ is
still matter of debate, ranging between 4 and 5. An important
characteristic of this relation is its extremely small scatter,
consistent with measurements errors only, so that equation (7) can be
considered a ``perfect'' relation.

In the context of galaxy formation studies, a natural question to ask
is how well {\it dissipationless} merging is able to produce and
maintain these relations. We focus on dissipationless merging since it
is simpler to be modeled with respect to gas rich merging, and because
in this way we can check whether the contribution of gas dissipation
is required in the merging scenario of Es formation.  In addition, the
analysis of dissipationless merging should indicate whether the
merging between gas poor systems observed at $z<1$ (van Dokkum et
al. 1999) is a common phenomenon in Es lifetime, or a rare event.
Among others, existence of serious problems encountered by the picture
of Es formation mainly driven by dissipationless merging to reproduce
their scaling relations was pointed out by Ciotti \& van Albada
(2001; hereafter CvA). In their {\it phenomenological approach}, just
by combining the observed (edge--on) FP and the $\Mbh$-$\sg0$
relations, they demonstrated that in a dissipationless merging
scenario, and under reasonable assumptions for the addition of BH
masses during the merging, {\it elliptical galaxies forced to lie on
the FP and to satisfy the $\Mbh$-$\sg0$ relation} would have effective
radii exceedingly larger than those observed in real Es.

Here we follow a complementary approach, and the main goal is
verifying, by using high resolution N-body numerical simulations of
one and two--component galaxy models, whether the end--products of
merging of galaxies, initially lying on the FP, lie on the FP (as well
as on the other scaling relations) or they fail in some other respect
(see also Pentericci, Ciotti \& Renzini 1995; Capelato, de Carvalho \&
Carlberg 1995; Bekki 1998; Evstigneeva, Reshetnikov \& Sotnikova 2002;
Nipoti, Londrillo \& Ciotti 2003a).

Numerical simulations are needed since, owing to possible structural
and/or dynamical non--homology and to projection effects, one cannot
predict with simple theoretical arguments how $\sg0$ and $\cRe$ evolve
as consequence of merging: however, rough indications on their
behavior can be derived from the study of the evolution of the {\it
virial velocity dispersion} $\sgv$ and the {\it virial radius}
$\rv$\footnote{By definition in a one--component galaxy $\sigma^2_V
\equiv 2T/M$ and $r_{\rm V} \equiv -GM^2/U$, where $T$ and $U$ are the
total kinetic and the gravitational energy of the galaxy,
respectively.}. In fact, as a consequence of the virial theorem and
the conservation of the total energy, in the merging of two galaxies
with masses $M_1$ and $M_2$ and virial velocity dispersions
$\sigma_{\rm V,1}$ and $\sigma_{\rm V,2}$, the virial velocity
dispersion of the resulting galaxy (in case of no mass loss and
negligible kinetic and interaction energies of the galaxy pair when
compared to their internal energies) is given by
\begin{equation}
\sigma^2_{\rm V,1+2}={M_1\sigma^2_{\rm V,1}+M_2\sigma^2_{\rm V,2} \over M_1+M_2}.
\end{equation}
  It follows that $\sigma_{\rm V,1+2}{\leq}\,{\rm max}(\sigma_{\rm
V,1},\sigma_{\rm V,2})$, i.e., {\it the virial velocity dispersion
cannot increase in a merging process of the kind described above}.
Under the same hypotheses, the virial radius $r_{\rm V,1+2}$ of the
resulting galaxy is given by
\begin{equation}
{(M_1+M_2)^2 \over r_{\rm V,1+2}}={M_1^2 \over r_{\rm V,1}} +{M_2^2 \over r_{\rm V,2}},
\end{equation}
where $r_{\rm V,1}$ and $r_{\rm V,2}$ are the virial radii of the
progenitors.  Identity (9) implies that \\$r_{\rm V,1+2} {\geq}\,{\rm
min}(r_{\rm V,1},r_{\rm V,2})$, i.e., {\it the virial radius cannot
decrease in a merging process of the kind described above}.

From equations (8) and (9) it follows that, in the highly idealized
scenario of a merging hierarchy based on identical, one--component
seed galaxies characterized by $\sgvzero$, $\rvzero$ and $M_0$, we
expect $\sgv=\sgvzero$ and $\rv=(M/M_0)\rvzero$, independently of the
merging sequence. Thus, by qualitatively assuming that $\sg0 \sim
\sgv$ and $\cRe \sim \ \rv$, one should conclude that the FJ and
Kormendy relations are not consistent with the scenario depicted
above.  Now, it is well known that for a large variety of mass models
there is a good correlation between $\rme$ (the half--mass radius of
the galaxy, strictly related to $\cRe$) and $\rv$, with the
proportionality constant showing little dependence on the particular
density profile\footnote{For example in truncated power--law spherical
models $\rv / \rme =2^{1 \over 3-\gm} (5-2\gm) / (3-\gm)$ and so $2
\leq \rv/\rme \lsim 2.1$ for $0\leq \gm \leq 2$.}  (see, e.g., Spitzer
1969, Ciotti 1991), and so, even in presence of structural
non--homology in the merging end--products, the ``prediction'' of a
linear growth of $\cRe$ with $M$ should be quite robust. A more
problematic situation (that can be properly addressed only with
numerical simulations) arises when considering the relation between
$\sgv$ and $\sg0$: at variance with $\rme/\rv$, structural
non--homology can {\it strongly} affect the ratio $\sg0/\sgv$ (see, e.g.,
Ciotti, Lanzoni \& Renzini 1996; Bertin, Ciotti \& Del Principe
2002). In addition, a systematic variation of orbital anisotropy with
mass (the so--called dynamical non--homology) can produce an
increasing $\sg0$ even at constant $\sgv$ (although Nipoti, Londrillo
\& Ciotti 2002, hereafter NLC02, showed that the whole tilt of the FP
cannot be ascribed to orbital anisotropy effects only).

Here, with the aid of high--resolution N-body simulations of one and
two--component galaxy models, we investigate the effect of subsequent
generations of merging on the galaxy scaling relations.  Our study
explore two extreme situations, namely the case of {\it major
merging}, in which equal mass galaxies are involved at each step of
the hierarchy, and the case of {\it accretion}, in which a massive
galaxy increases its mass by incorporating smaller galaxies: the
evolution of more realistic, ``mixed'' merging histories should be
bracketed by our simulations.  This paper is organized as follows.  In
Section 2 we describe the properties of the galaxy models and the
adopted numerical methods.  In Section 3 we present the results, while
in Section 4 we specifically discuss the consequences of the findings
on the implications for the $\Mbh$-$\sg0$ relation.  Finally, in
Section 5 the main conclusions are summarized.

\section{Numerical methods}

\subsection{Galaxy models}
As initial conditions for the first generation of merging we use
spherically symmetric one and two--component Hernquist density
distributions (Hernquist 1990; Ciotti 1996, 1999). This choice is
motivated by the fact that the Hernquist model is both a reasonable
approximation (when projected) of the $R^{1/4}$ law (de Vaucouleurs
1948), and it is also similar to the Navarro, Frenk \& White (1996)
profile, thus providing a sufficiently realistic description of the
galaxy stellar and dark matter (DM) density distributions.  The
density, mass, and (relative) potential profiles of the stellar
component of the Hernquist model are given by
\begin{equation}
\rhostar (r)= {1 \over 2\pi}{\Mstar\rc \over r (\rc+r)^3},
\end{equation}
\begin{equation}
{\Mstar (r)\over\Mstar}=\left({r\over \rc+r }\right)^2,
\end{equation}
\begin{equation}
{\Psistar (r)}={G\Mstar \over \rc +r },
\end{equation}
where $\Mstar$ is the total stellar mass. In the two--component
models, the DM halo is described by $\rho_{\rm h}$, $M_{\rm h}$ and
$\Psi_{\rm h}$ profiles of the same family of equations (10)-(12),
where now $r_{\rm h} \equiv {\beta} \rc$ and $M_{\rm h} \equiv {\mu}
\Mstar $. At variance with the models used in NLC02, here we restrict
to globally isotropic models: the distribution function (DF) of the
stellar component is then given by
\begin{equation}
f_*(\en)=\frac{1}{\sqrt{8}\pi^2}\frac{d}{d\en}
       \int_0^\en{\frac{d{\rho_*}}{d\psit}}{\frac{d\psit}{\sqrt{\en-\psit}}},
\end{equation}
where $\en =\psit-v^2/2$ is the relative (positive) energy, $v$ is the
modulus of the velocity vector, and $\psit=\Psistar +\Psi_{\rm h}$ is
the relative total potential; an analogous identity holds for the DM halo, where $\rho_{\rm h}$ substitutes $\rho_*$. 

According to the given definitions, the one--component models are
completely determined by the two physical scales $\Mstar$ and $\rc$,
and in the numerical simulations $\Mstar$, $\rc$ and $\Td$ are adopted
as mass, length and time scales.  In particular $\Td$ is the
half--mass dynamical time, defined as
\begin{equation}
\Td \equiv \sqrt{\frac{3\pi}{16G\rhoM}} \simeq
           8.33\sqrt{\frac{\rc^3}{G\Mstar}},
\end{equation}
where $\rhoM =3\Mstar /8\pi \rme^3$ is the mean density inside the
half--mass radius $\rme$. Finally, the velocity scale is given by
\begin{equation}
v_{\rm c}\equiv\frac{\rc}{\Td} \simeq {0.12}\sqrt{\frac{G\Mstar}{\rc}}.
\end{equation}
The two--component Hernquist models are characterized by four
quantities, the two physical scales $\Mstar$ and $\rc$, and the two
dimensionless parameters $\mu$ and $\beta$.  In order to transform the
results of the numerical simulations in physical units we express
$\Mstar$ in $10^{10}\Msol$, $\rc$ in kpc, $\Lb$ in $10^{10}\Lbsol$,
and the constant {\it stellar} mass--to--light ratio $\ml \equiv
M_*/\Lb$ in solar units. For a detailed description of the numerical
realization of the initial conditions, see Section 2.1.3 of NLC02.

\subsection{Initial conditions}

\subsubsection{Equal mass merging}

The first generation of the {\it equal mass merging} hierarchy is
obtained by merging a pair of identical, spherically symmetric and
isotropic Hernquist models (the ``zeroth order'' seed galaxies), while
the successive generations are obtained by merging pairs of identical
systems obtained by duplicating the end--product of the previous
step. We followed the evolution of the hierarchy of 5 steps of {\it
head--on} mergers (that would correspond, in absence of mass escape,
to a mass increase of a factor 32); we also explored 3 steps of the
{\it head--on} merging hierarchy whose seed galaxies are
two--component Hernquist models (in which we assume a DM halo more
massive, $\mu=3$, and less concentrated, $\beta=2$, than the stellar
component), and, finally, 3 steps of the hierarchy of encounters of
one--component models with {\it non zero (orbital) angular momentum}.

In the assignment of the initial conditions for the head--on
encounters the two galaxies are initially placed on an orbit
characterized by vanishing relative energy and angular momentum, i.e.,
they would have a null relative velocity while at infinite relative
distance.  In practice, at the time $t=0$, we place the two galaxies
at a distance $d_{\rm rel}\simeq3 {\r90}$ (where $\r90$ is the
angle--averaged radius enclosing the $90\%$ of the total mass $M$ of
each galaxy), and, neglecting the effects due to tidal forces, we
assign to their centers of mass a relative velocity with modulus
$v_{\rm rel}=2\sqrt{GM/d_{\rm rel}}$.  The end--products of each
merging are non--spherical, and so, when exploring a successive
merging, their mutual orientation at the beginning of the new
simulation is randomly assigned. In the simulations with orbital
angular momentum, we still consider vanishing relative energy and
$d_{\rm rel}$ defined as above, and we assume as impact parameter the
sum of the angle--averaged half--mass radii of the merging galaxies.

\subsubsection{Accretion}

In the case of the merging hierarchy in which the test galaxy grows by
accretion of smaller systems ({\it accretion}), the seed galaxy is a
one-component Hernquist model, and the first merging event is
identical to that in the equal mass merging case. In the second step,
however, the first end--product (of mass $\sim 2M_*$) merges again
with a seed galaxy, and so on. As a consequence, the mass ratio
between the infalling stellar system and the test galaxy decreases
approximately as $1/n$, where $n$ is the step in the accretion
hierarchy. Due to the higher computational cost required by accretion
simulations with respect to equal mass merging in order to reach the
same mass increase, in the former case we limit to 15 steps (of
head--on merging) and to 9 steps (of merging with angular momentum),
for a putative mass increase of a factor 16 and 10, respectively.  As
in the case of equal mass merging simulations, the relative initial
positions and velocities of the infalling satellite and of the test
galaxy are chosen so that at $t=0$ they are on parabolic orbit, and
the mutual orientation of the two galaxies is assigned randomly. In
practice, as initial relative distance we adopt $d_{\rm rel}\simeq
\runo+2\rdue$, where the subscript 2 refers to the smaller
galaxy. Again, in case of non zero angular momentum, the impact
parameter is given by the sum of the angle--averaged half--mass radii.

\subsection{The numerical code}

For the simulations we used a new, fast and accurate, parallel N-body
code. Recently, a numerical scheme has been introduced to simulate
collisionless N-body systems by Dehnen (2000, 2002), with a
substantial improvement in force computation over standard tree--based
codes. By combining in an original way the Fast Multipole Algorithm
(see, e.g., Greengard \& Rokhlin 1987, 1997) with the Barnes \& Hut
(1986) tree data structure, Dehnen scheme achieves exact momentum
conservation with an effective $O(N)$ operational complexity (for particle numbers $N \gsim 10^5$).

By taking advantage of these improvements, we have implemented the
Dehnen fast force solver (FalcON) in a parallel Fortran-90 N-body code
named FVFPS (Fortran Version of a Fast Poisson Solver; Londrillo,
Nipoti \& Ciotti 2003). The FVFPS code is based on domain
decomposition, and uses the MPI routines for data communication. A
mass dependent opening parameter $\theta=\theta(M)$, that assures
faster performances and a more uniform error distribution, is also
employed (Dehnen 2002). Time integration is performed by a standard
leap-frog algorithm, with (uniform) time step $\Dt$ determined
adaptively. Thanks to effective $O(N)$ scaling of the force
computations, the FVFPS code assures a speed-up of a factor $\sim 10$
over existing codes, in particular over the Springel, Yoshida \& White
(2001) GADGET code.  In the present version, the FVFPS code is
characterized by three preassigned parameters:

{\it (i)} The minimum value of the opening parameter $\thetamin$,
associated with the total mass. We adopt $\thetamin=0.5$, resulting in
mean relative force errors of $\sim 0.3 \%$, and 99 percentile errors
of $\sim 3 \%$.

{\it (ii)} The softening parameter $\varepsilon$ (i.e., the softening
length expressed in units of $\rc$), dependent, according to
literature indications (e.g., Merritt 1996, Athanassoula et al. 2000,
Dehnen 2001), on $N$ and on the specific density distribution profile.

{\it (iii)} For the initial time step we assume $\Dt \sim \Td/100$.
The time step is allowed to vary adaptively as a function of the
maximum particle density, but is kept the same for all the particles.

For the stellar distribution of the seed galaxies we use as a rule
$N_*=16378$ particles for each galaxy, and in the two--component cases
the DM distribution is made of $N_{\rm DM}=49152$ particles: with this
choice, halo and stellar particles have the same mass. As a
consequence of the merging hierarchy, the number of particles in the
simulations increases with the galaxy mass: in the final merging of
(one and two--component) equal mass systems and of the accretion
scenario, the total number of particles involved is of the order of
$5.2 \times 10^5$ and $2.6 \times 10^5$, respectively.

We followed the dynamical evolution of each merging event up to the
virialization of the resulting system, which is usually reached on a
time scale shorter than $50\Td$ after the first encounter between the
two galaxies: for example the virialization time for the first step in
the one--component (head--on) equal mass merging hierarchy is of the
order of $2\,{\rm Gyr}$ having assumed $M_*=10^{10}\Msol$ and
$\rc=1\,{\rm kpc}$. In the following we call ``end--product'' of a
simulation the system made of bound particles after the virialization.
For the determination of the intrinsic and ``observational''
properties of the end--products, we followed the same procedure
described in NLC02. In particular we measured the end--product
intrinsic axis ratios $c/a$ and $b/a$ (where $a$, $b$ and $c$ are the
major, intermediate and minor axis of the associated inertia
ellipsoid), their virial velocity dispersion $\sgv$ and
angle--averaged half--mass radius $\rme$, and, for several projection
angles, the end-product circularized effective radius $\cRe$, the
central velocity dispersion $\sg0$, and the mean effective surface
brightness $\Ie$ (obtained from the stellar mass profile by assuming a
constant stellar mass--to--light ratio $\ml$).  In particular, the
central velocity dispersion $\sg0$ is obtained by averaging the
projected velocity dispersion over the circularized surface brightness
profile within a radius of $\cRe/8$. As a general rule we find that
discreteness effects on the derived values of $\cRe$ and $\sg0$ do not
exceed $1\div 2\%$ at each projection angle.

\begin{figure}
\begin{center}
\parbox{1cm}{ \psfig{file=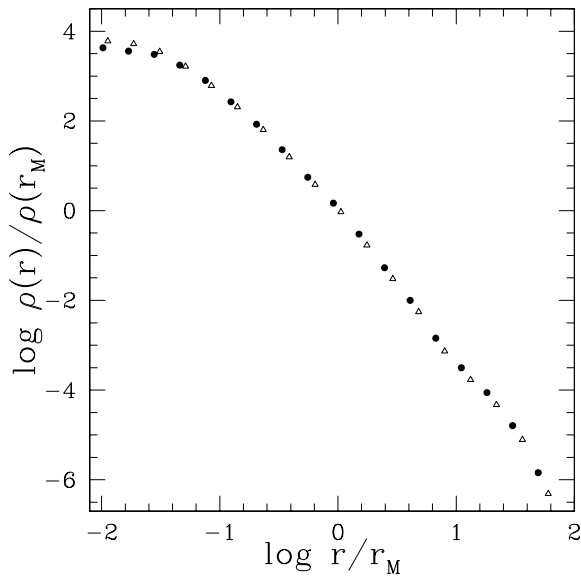,width=0.45\textwidth}}
\caption{Angle-averaged density as a function of radius (normalized to
the half--mass radius $\rme$) of the end--product of the head--on
merging of two one--component Hernquist models, obtained with GADGET 
(empty triangles) and with the FVFPS code (solid circles). }
\end{center}
\end{figure}

We checked the reliability of the numerical simulations by running a
few merging events both with the parallel version of GADGET (with
$\alpha=0.02$, $\Dtmin =0$, $\Dtmax=\Td/100$, $\alphatol =0.05$, and
$\varepsilon =0.05$ ;see Springel et al. 2001 for details) and with
our FVFPS code (with $\thetamin=0.5$, $\varepsilon =0.05$ and $\Dt
=\Td/100$). The results are in remarkable agreement: the end--products
in three-dimensional shape, angle--averaged half--mass radius, density
and velocity dispersion profiles are practically indistinguishable.
With both codes the total energy is conserved within $1\%$ over 100
dynamical times. As an illustrative example, in Fig. 1 we show the
density profiles of the end--products of the head--on merging of two
Hernquist models (GADGET, empty triangles; FVFPS, solid circles); note
how ``minor'' details, as the slope change in the density profile at
$r \sim 10 \rme$, are nicely reproduced. All the simulations were run
on a Cray T3E, and on a IBM Linux Cluster (with a number of processors
from 4 up to 32, depending on the number of particles of the
simulation).

\section{The results}

\subsection{Equal mass merging}

Before presenting the effects of equal mass merging on the FP and on
the other scaling relations, we briefly describe the induced
modifications of the internal structure and dynamics of one and
two--component galaxy models. In general, the end--products are
triaxial systems with axis ratios in the range $0.5\, \lsim \, c/a \,
\lsim \,0.7$ and $0.7\, \lsim \, b/a \, \lsim \,0.8$ (where $a$, $b$
and $c$ are the major, intermediate and minor axis, respectively), in
accordance with observed ellipticities of Es.  For comparison with
real galaxies, we also fitted (over the radial range $0.1 \, \lsim \,
R/\cRe \, \lsim \, 10$) the projected stellar mass density profile of
the end--products with the Sersic (1968) $R^{1/m}$ law:
\begin{equation}
I(R)=I_0\,\exp\left[-b(m)\left(\frac{R}{\Re}\right)^{1/m}\right],
\end{equation}
where $b(m)\simeq 2m-1/3+4/405m$ (Ciotti \& Bertin 1999), and 
\begin{figure}
\begin{center}
\parbox{1cm}{ \psfig{file=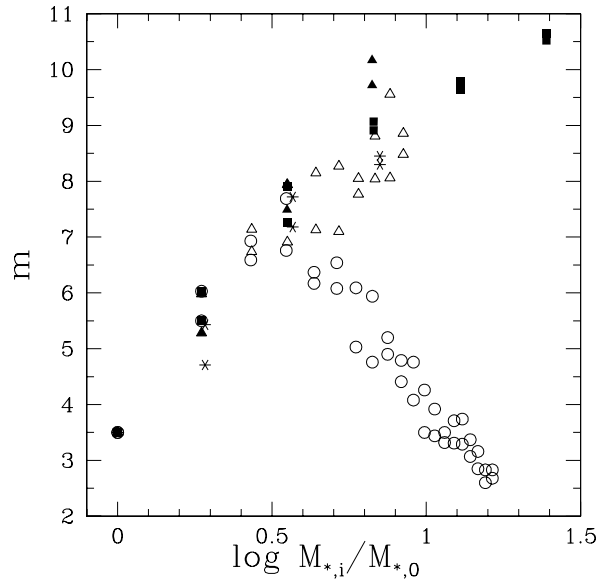,width=0.45\textwidth}}
\caption{Sersic best fit parameter $m$ vs. total stellar mass of the
end--products at stage $i$ of the merging hierarchy. Equal mass
mergers are shown as solid triangles and squares (one--component
galaxies), and stars (two--component galaxies); empty triangles and
circles represent the accretion hierarchies. Triangles correspond to
simulations with non zero orbital angular momentum.}
\end{center}
\end{figure}
in Fig. 2 we plot the best fit parameter $m$ as a function of the
total mass of the systems, for one--component (solid squares and
triangles) and two--component (stars) models. Clearly, the fitted
quantities $m$ and $\cRe$ depend on the relative orientation of the
line--of--sight and of the end--products of the simulations: the two
points for each value of the mass in Fig. 2 show the range of values
spanned by $m$ when projecting the final states along the shortest and
longest axis of their inertia ellipsoids.  We note that higher values
of the best fit parameter $m$ correspond to more massive systems, a
trend similar to that observed in Es\footnote{Curiously, NLC02 found
that in the case of end--products of unstable galaxy models $m$
decreases for increasing radial orbital anisotropy in the initial
conditions. Note also that in NLC02 and Londrillo et al. 2003 we used
a radial range $0.1 \, \lsim \, R/\cRe \, \lsim \, 4$.} (see, e.g.,
Caon, Capaccioli \& D'Onofrio 1993, Prugniel \& Simien 1997, Bertin et
al. 2002). Note also that the values of $m$ are in the same range
obtained from observations, in fact we found $2\,\lsim\, m\,\lsim\,
11$. (However, as pointed out to us by the Referee, from Fig. 2 one
could argue that $m \approx 4$ galaxies should have experienced at
most one major merger event in their life.)

Following the discussion in the Introduction, in Fig. 3 (top panel) we
show the relation between the virial velocity dispersion and the total
stellar mass of the mergers at each step of the merging hierarchy. In
one--component models (solid squares and triangles) a modest increase
of $\sgv$ with $M_*$ is apparent: for the last model in the hierarchy
the mass is increased by a factor $\sim 24.5$ (instead of the maximum
possible value of 32), while the $\sgv$ is increased by a factor $\sim
1.16$ only. This small increase of $\sgv$ with respect to the
expectation of equation (8) (i.e., $\sgv=const$) can be explained with
a simple generalization of the highly idealized situation formalized
there.  In fact, if during a merging between two galaxies with mass
$M_1$, $M_2$ and virial velocity dispersion $\sigma_{\rm V,1}$,
$\sigma_{\rm V,2}$ a mass $\Delta M$ is lost with mean velocity
$v_{\rm ej}$, then the virial velocity dispersion of the end--product
$\sigma_{\rm V,1+2}$ is expected to be
\begin{equation}
\sigma^2_{\rm V,1+2}={M_1\sigma^2_{\rm V,1}+M_2\sigma^2_{\rm V,2}+\Delta Mv_{\rm ej}^2 \over M_1+M_2-\Delta M}:
\end{equation}
in particular $\sgv$ increases as a consequence of the mass (and
associated kinetic energy) loss.  We found that in the simulations the
mass lost during each merging never exceeds $\Delta
M/(M_1+M_2)\simeq0.06$ and, consistently, the corresponding increase
of $\sgv$ is $2\div5\%$ (Fig. 3, top panel).

\begin{figure*}
\begin{center}
\parbox{1cm}{ \psfig{file=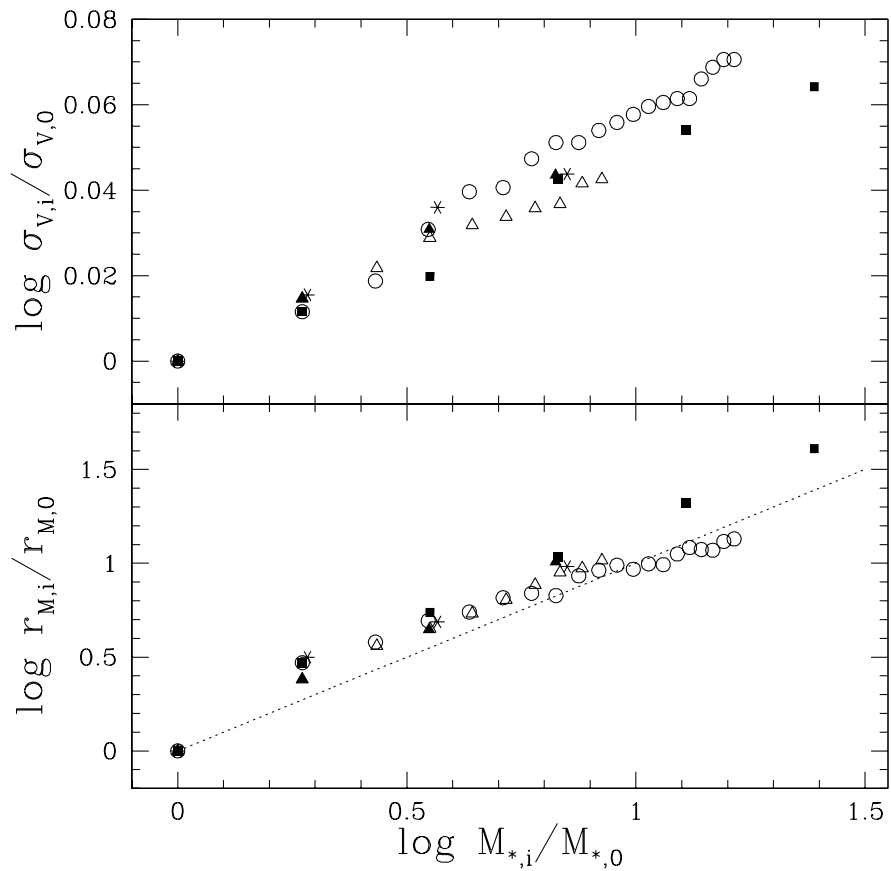,height=0.5\textheight}}
\caption{{\it Top panel}: virial velocity dispersion of the stellar
component at stage $i$ of the merging hierarchy vs. the total stellar
mass of the merger. {\it Bottom panel}: angle--averaged half--mass
radius $\rme$ vs. total stellar mass. Symbols are the same as in
Fig. 2. The dotted line indicates the relation $\rme \propto M$;
$M_{*,0}$, $r_{\rm M,0}$ and $\sigma_{\rm V,0}$ are the stellar mass,
the virial velocity dispersion and the half--mass radius of the seed
galaxy. In the two--component cases, $\sgv$ and $\rme$ refer to the
stellar component only. Note the different range spanned in the
ordinate axes in the two panels.}
\end{center}
\end{figure*}

In Fig. 3 (bottom panel) we plot the half--mass radius as a function
of the stellar mass of the mergers. As for $\sgv$, also for this
quantity the simple virial expectation of a linear growth of $\rv$
(and of $\rme$) with $M$ (dotted line) is apparent. The ``jump'' from
the initial condition position to the first merger is due to the
significant change in the galaxy density structure (as revealed by the
change of values of $m$ in Fig. 2); the successive generations of
merging are characterized by more similar density profiles, and,
correspondingly, the points in Fig. 3 move parallel to the dotted
line.

How does the presence of massive DM halos affect these results? As
anticipated, in this case we investigated 3 steps of the (head--on)
equal mass merging hierarchy by using two--component galaxy
models. Due to its observational relevance, we focus here on the
description of the properties of the stellar component only. As in the
one--component case, the end--products are triaxial, and with the axis
ratios in the same range. The best fit Sersic parameter $m$ still
increases with mass (stars in Fig. 2).  In the two--component
simulations, owing to the presence of the DM halo, there are not
arguments as simple as those used in equations (8), (9) and (17) in
order to predict the effects of merging on $\sgv$ (defined as
$2T_*/M_*$) and $\rme$ of the {\it stellar} component. However, it is
apparent from Figs. 2 and 3 that the presence of a DM halo or angular
momentum in the initial conditions does not modify the general trend
of $m$, $\sgv$ and $\rme$ at increasing mass.

\subsubsection{Fundamental Plane}

\begin{figure*}
\begin{center}
\parbox{1cm}{ \psfig{file=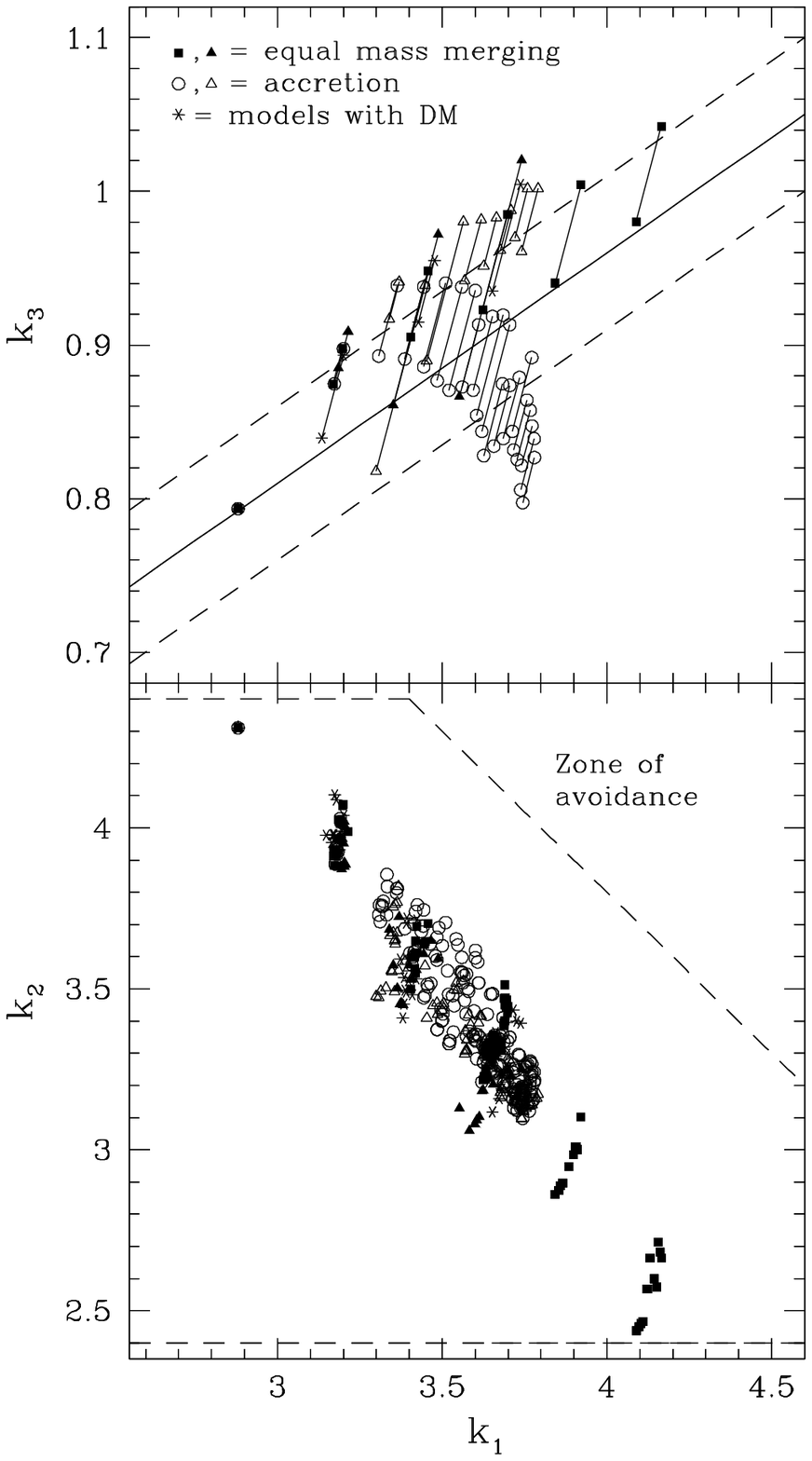,height=0.9\textheight}}
\caption{ {\it Top panel}: the merging end--products in the
($\ku,\kt$) plane, where the solid line represents the FP relation as
given by equation (4) with its observed 1-$\sigma$ dispersion (dashed
lines). Bars show the amount of projection effects. {\it Bottom
panel}: the merging end--products in the ($\ku,\kd$) plane, where the
dashed lines define the region populated by real galaxies as given by
BBF. Each model is represented by a set of points corresponding to
several random projections. Symbols are the same as in Fig. 2.}
\end{center}
\end{figure*}

In Fig. 4 we plot the results of equal mass merging simulations in the
principal planes $(\ku,\kt)$ and $(\ku,\kd)$, where one--component
galaxy models resulting from head--on mergers and from merging with
angular momentum are identified by solid squares and triangles,
respectively; two--component models are represented by stars.  The
progenitor of the merging hierarchy (the point without bar) is placed
on the edge--on FP by choosing its luminosity $\Lb = 3 \times 10^9
\Lbsol$ and {\it stellar} mass--to--light ratio $\ml=5$, so that
equation (4), represented by the solid line, is satisfied.
Consistently with the adopted dissipationless scenario, the value of
$\ml$ is kept constant during the whole merging hierarchy. Due to the
loss of spherical symmetry of the end--products of the merging
simulations, their coordinates depend on the line--of--sight
direction; however, being the luminosity (mass) of each end--product
fixed, variations of $\ku$, $\kd$, $\kt$ due to projection effects are
not independent. In fact, the $\ku$ and $\kt$ coordinate of a galaxy
of given luminosity are linearly dependent as
\begin{equation}
\kt=\sqrt{\frac{2}{3}}\ku+{\sqrt{\frac{1}{3}}}\log {\frac{2\pi}{\Lb}};
\end{equation} 
this dependence is reflected by the straight lines in Fig. 4 (top
panel), each of them representing the range spanned in the ($\ku,\kt$)
space by a given end--product when observed over the solid angle. As
in NLC02, in order to quantify the deviation of a model from the FP,
we use the vertical distance $\delta\kt\equiv |\kt -0.15\ku -0.36|$
from the point to the FP itself. It is interesting to note that for
all the end--products the projection effects translate into a
$\delta\kt$ of the same order of magnitude of the observed FP
dispersion.  In addition it is also apparent from Fig. 4 how $\kt$ and
$\ku$ increase with the galaxy mass consistently with the observed FP
tilt and thickness. Being $\ml$ fixed in our simulations, from Fig. 2
we conclude that one--component equal mass dissipationless merging is
able to reproduce (basically by structural non--homology) the edge--on
FP of elliptical galaxies (see also Capelato et al. 1995).

We also explored the behavior of the models in the $(\ku,\kd)$ plane
(Fig. 4, bottom panel), which represents the face--on view of the FP;
the region populated by real galaxies in this plane is identified by
dashed lines.  By fixing the scale length, we place the first
progenitor (with $\cRe\simeq0.7\,{\rm kpc}$) in a zone of the
$(\ku,\kd)$ plane populated by low--luminosity ellipticals
($\ku\simeq3$, $\kd\simeq4.5$), and for each end--product we plot the
($\ku,\kt$) positions for a few random projection angles. Note that,
at variance with the coordinates ($\ku,\kt$), $\ku$ and $\kd$ are
related (for a given end--product) by an expression containing $\cRe$
besides $\Lb$, and for this reason in the $(\ku,\kd)$ plane the effect
of projection is to distribute the end--products on two--dimensional
regions.  One--component equal mass mergers (solid squares and
triangles) are moved by merging towards the bottom right of the
($\ku,\kd$) plane, roughly parallel to the line defining the {\it zone
of avoidance}, in accordance with the prediction of BBF.  The produced
displacements are very large: it is remarkable that the end--products
of the last step of the hierarchy are found in a position only
marginally consistent with the populated region in the ($\ku$,$\kd$)
space. It is also interesting to note that the presence of substantial
angular momentum in the initial conditions does not change
significantly the properties of equal mass mergers, both in the
($\ku$,$\kt$) and in the ($\ku$,$\kd$) plane.

The results of equal mass merging of two--component galaxies are
represented in Fig. 4 with stars. Due to the presence of DM, in order
to place the first progenitor ($\Lb= 3 \times 10^9 \Lbsol$) on the FP,
we assumed a stellar mass--to--light ratio $\ml=2.3$. It is apparent
that both in the ($\ku,\kt$) and ($\ku,\kd$) planes, the behavior of
two--component models does not significantly differ from the
corresponding one--component models. In other words, {\it
dissipationless, equal mass merging of one and two--component models
seems to be consistent with the existence of the FP of Es}.

\subsubsection{Faber--Jackson and Kormendy relations}

\begin{figure*}
\begin{center}
\parbox{1cm}{ \psfig{file=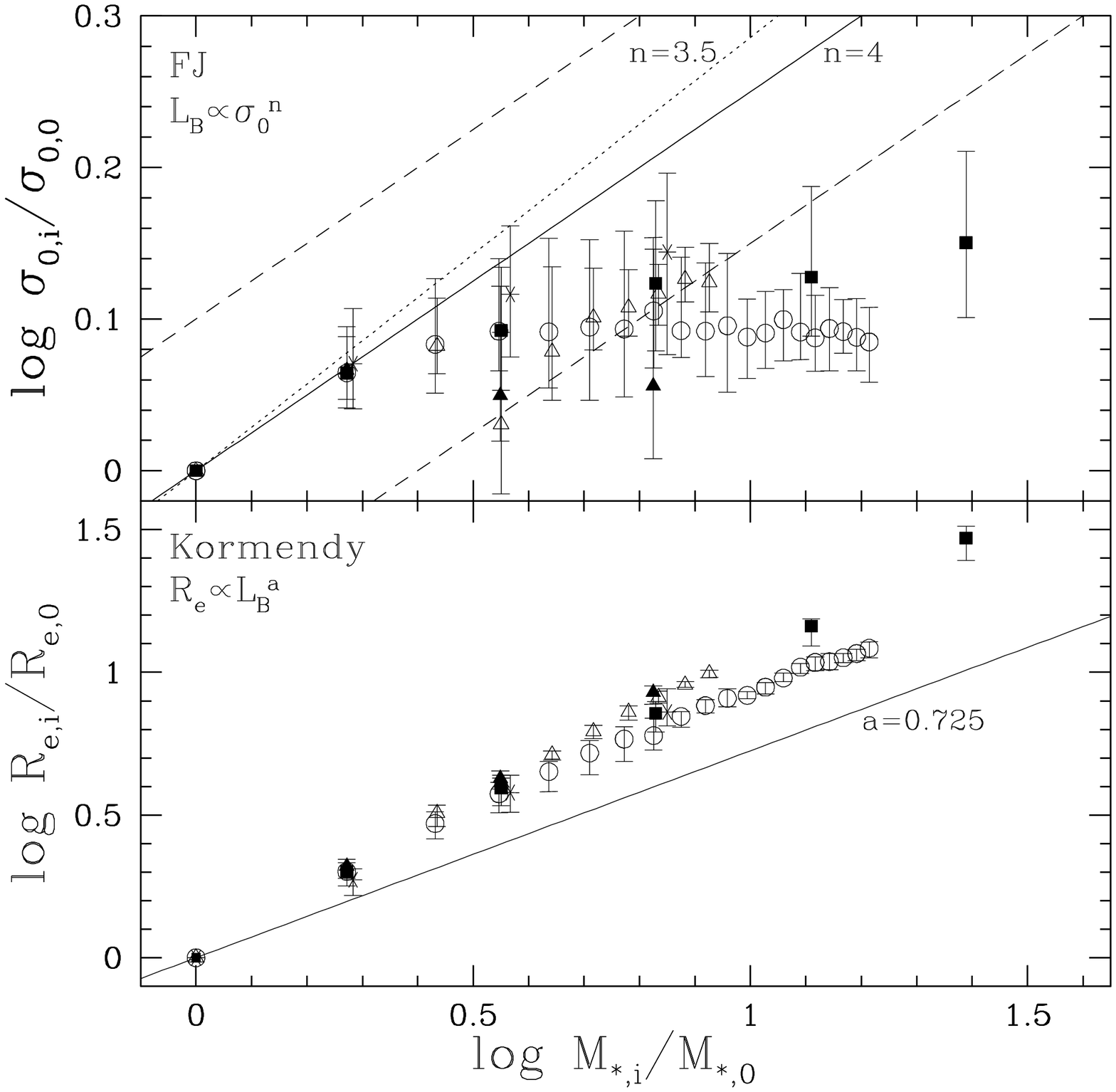,height=0.5\textheight}}
\caption{{\it Top panel}: stellar central velocity dispersion
(normalized to that of the first progenitor) vs. total stellar
mass. Points correspond to angle--averaged values, bars indicate the
range spanned by projection effects.  The solid and the two dashed
lines represent the Faber-Jackson relation $\Lb \propto \sg0^4$ and
its scatter, while the dotted line represents $\Lb \propto
\sg0^{3.5}$. {\it Bottom panel}: stellar effective radius (normalized
to that of the first progenitor) vs.  total stellar mass. Points and
bars have the same meaning as in top panel. The solid line represents
our ``fiducial'' Kormendy relation, obtained combining the FP and the
FJ with $n=4$ (see text); symbols are the same as in Fig. 2. Note the
different range spanned in the ordinate axes in the two panels. }
\end{center}
\end{figure*}

As we have shown in the Section above, dissipationless merging of one
and two--component galaxy models is surprisingly consistent with the
existence of the FP, especially considering its small thickness when
seen edge--on. However, Es do follow additional scaling relations as
the FJ and Kormendy relations, and so here we compare the results of
our simulations with these scaling laws.

The solid line in the upper panel of Fig. 5 represents the FJ relation
in the form $\Lb\propto \sg0^4$ together with its scatter ($\delta
{\rm log}\sg0\simeq0.1$, Davies et al. 1983, dashed lines); the dotted
line represents the steeper fit ($\Lb\propto \sg0^{3.5}$) derived by
Dressler et al. (1987). The end--products of (head--on) equal mass
merging of one--component galaxies (solid squares) have a $\sg0$ lower
than that predicted by the FJ relation for the given mass increase,
and the discrepancy is stronger in case of merging with angular
momentum (solid triangles). It is interesting to compare this plot
with the upper panel of Fig. 3: it is then clear how the failure of FJ
can be directly traced to the constancy of $\sgv$ {\it and} to the
significant dynamical homology of the mergers. This latter point again
suggest that {\it structural} non--homology, more than dynamical
non--homology is at the origin of the FP in the explored merging
scenario (see also NLC02).  Models with DM (stars in Fig. 5) seem to
follow more closely the FJ relation, even though the representative
points show the same clear trend as the one--component merging
hierarchy: it seems clear to us that additional simulations would
bring also two--component models completely outside the FJ relation.

Considering now the agreement with the FP on one side, and the failure
at reproducing the FJ relation on the other side, we would expect that
also the Kormendy relation is not reproduced by the equal mass merging
hierarchy. In fact, if we assume that the Es obey to the FJ relation,
$\Lb \propto \sigma^n$, then from equation (4) we obtain a fiducial
Kormendy relation $\cRe \propto \Lb^{1.225-2/n}$. For example, with
$n=4$ (for which the exponent is remarkably near to that reported by
Bernardi et al. 2003) the effective radius of the end--product of the
last equal mass merger is a factor $\sim 3$ larger than that predicted
by the ``Kormendy'' relation. In other words, the results of our
simulations are characterized by too large effective radii (see
Fig. 5, bottom panel).  Thus, in the explored equal mass merging
hierarchies both $\sg0$ and $\cRe$ deviate systematically from the
values expected from FJ and Kormendy relations.  Yet, their deviations
from the two scaling laws curiously compensate reproducing the
edge--on FP; this shows clearly that the FJ and Kormendy relations are
not simple projections of the FP, which is actually preserved by
merging, but, despite their larger scatter, contain additional
information on galaxy structure and dynamics.
 
\subsection{Accretion}

In Section 3.1 we showed that, in the case of equal mass merging of
one and two--component galaxies, the virial expectations were
qualitatively followed also by the observational properties $\cRe$ and
$\sg0$, in addition to $\sgv$ and $\rv$. We recall that these
expectations are independent of the specific merging history, as far
as the zeroth order seed galaxies are identical systems, and so we
could consider ``accretion'' simulations superfluous.  However, the
amount of escaping particles or the non--homology of the final
end--products could depend on the mass ratio of the merging systems,
thus resulting in substantial changes in $\sgv$ and/or
$\sg0$. Following this argument we now present the results of
accretion simulations, whose technical setting is described in Section
2. From the structural point of view, the end--products of accretion
are less flattened than the corresponding (i.e. with the same mass)
equal mass mergers, being, in general, triaxial systems with axis
ratios in the range $0.6\, \lsim \, c/a \, \lsim \,0.9$, and $0.7\,
\lsim \, b/a \, \lsim \,0.9$, and becoming more and more spherical
with increasing mass as a consequence of the higher number of merging
events and the random direction of accretion.  The evolution of the
surface brightness profile in the accretion hierarchy is shown in
Fig. 2, where the best fit Sersic parameter $m$ is represented with
empty circles and triangles (for head--on merging and merging with
angular momentum, respectively).  {\it A first relevant difference
with the equal mass merging results is the {\it decrease} of $m$ with
mass at mass ratios larger than 4 for the head--on accretions}, and
the consequent smaller range of $m$ ($2 \, \lsim \, m \, \lsim \, 8$)
spanned by their end--products. Therefore, the explored head--on
accretion scenario fails at reproducing the relation between the
surface brightness profile shapes and luminosity of real galaxies,
which are characterized by $m$ increasing with galaxy luminosity and
also spanning a larger range.  A different situation is obtained when
considering accretion simulations with non negligible orbital angular
momentum: the structural properties of the end--products are curiously
similar to the equal mass cases.

In the upper panel of Fig. 3 we plot $\sgv$ as a function of mass of
also of the head--on accretion end--products (empty circles). We found
that the fraction of mass lost in each merging $\Delta M/(M_1+M_2)$,
decreases from $\sim0.06$ in the first step to $\sim0.01$ in the last
one, summing up to a total amount of mass lost over 15 steps of $\sim$
$18\%$, about the same value found in the corresponding equal mass
merging cases.  The slightly larger increase of $\sgv$ in the head--on
accretion hierarchy with respect to the equal mass merging cases is
fully accounted for larger values of $v_{\rm ej}$, in accordance with
the estimate given in equation (17).  In the lower panel of Fig. 3 we
also show the evolution of $\rme$ in the head--on accretion hierarchy
(empty circles). The behavior of this quantity is qualitatively
similar to that of equal mass mergers: the slightly flatter slope of
the sequence with respect to the simple expectation $\rme \propto M$
is due to the stronger evolution of structural non--homology, as
revealed by Fig. 2. On the other hand, the end--products of accretion
simulations with angular momentum (empty triangles in Fig. 3) are more
similar to equal mass merging case.

\subsubsection{Fundamental Plane}

The most striking difference of the end--products of head--on
accretion simulations (empty circles) with respect to the equal mass
merging hierarchy and also to accretion simulations with angular
momentum (empty triangles) is apparent in the upper panel of Fig. 4,
where the results are plotted in the $(\ku,\kt)$ space. In fact, after
few accretions, the end--products are characterized by a $\kt$ {\it
decreasing} for increasing $\ku$, at variance with the FP slope and
the trend shown by the end--products of equal mass mergers.  As a
consequence, the last explored models (corresponding to an effective
mass increase of a factor $\sim 12$) are found at a distance $\delta
\kt$ larger than the FP scatter. This result, when interpreted by
using the information in Fig. 2, is not surprising. In fact, being the
coordinate $\kt$ a measure of non--homology for galaxies with constant
$\ml$ (as those explored in this paper), the decrease of the Sersic
parameter $m$ at increasing mass (at variance with real galaxies)
reflects directly in the unrealistic trend in the $(\ku,\kt)$
plane. On the contrary, in the face--on $(\ku,\kd)$ plane (Fig. 4,
lower panel), the end--products of accretion (both head--on and with
angular momentum) evolve along the same direction followed by equal
mass mergers, albeit with a smaller excursion in $\ku$ and $\kd$.

\subsubsection{Faber--Jackson and Kormendy relations}

In Fig. 5 we also plot the results of accretion simulations in the
planes representing the FJ and the Kormendy relations. At variance
with the case of the edge--on FP, and {\it independently on the
presence of angular momentum}, the behavior of the end--products is
considerably similar to that of the equal mass mergings, i.e., both FJ
and Kormendy relations are {\it not} reproduced. The end--products
have $\sg0$ smaller and $\cRe$ larger than the values predicted by the
two scaling relations for the given luminosity (mass) increase.
However, a closer inspection of Fig. 5 reveals that both $\sg0$ and
$\cRe$ are systematically larger in the case equal mass mergers than
in the case of head--on accretion: in other words, the end--products
in the latter case deviate {\it more} from the FJ and {\it less} from
the Kormendy than those in the former. The result relative to the
effective radius simply reflects the different evolution of the
half--mass radius in the two scenarios (see Fig. 3, lower panel),
while the behavior of $\sg0$ depends both on structural and dynamical
non--homology effects: the combined effect of the lower values of
$\sg0$ and the larger values of $\cRe$ is responsible of the
decreasing of $\kt$ with $\ku$ observed in the analysis of the
edge--on FP. In other words, at variance with equal mass mergers and
accretions with angular momentum, there is not enough compensation
between the trends in $\sg0$ and $\cRe$ able to reproduce the FP tilt
in case of head--on accretions.

\section{Dissipationless merging and the $\Mbh$-$\sg0$ relation}

\begin{figure*}
\begin{center}
\parbox{1cm}{ 
\psfig{figure=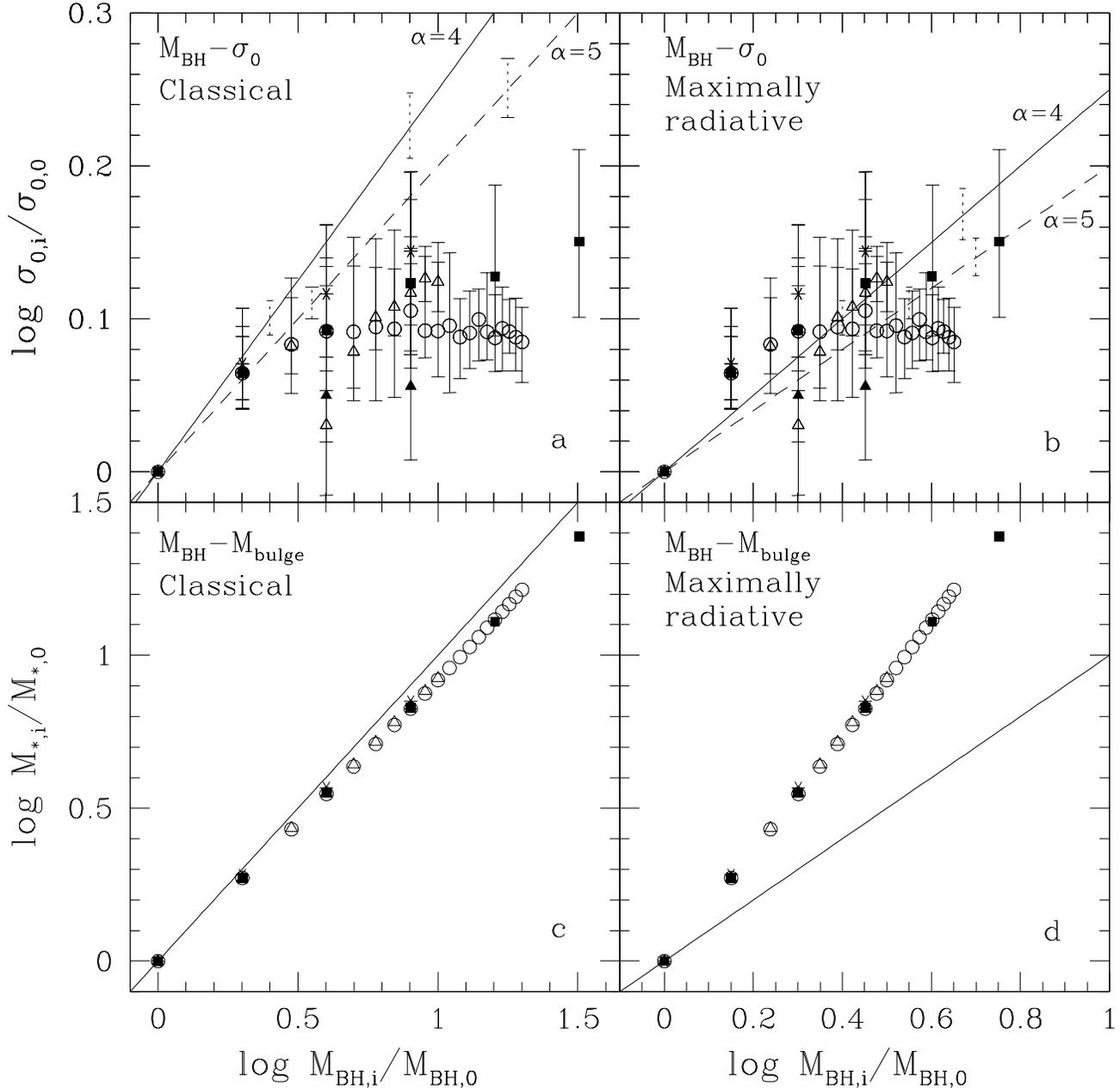,width=\textwidth}}

\caption{ {\it Panel a}: galactic central velocity dispersion vs.
BH mass for classical BH merging; $\sigma_{0,0}$ and $M_{\rm BH,0}$
are the central velocity dispersion and BH mass of the first
progenitor, respectively. The points correspond to the mean value over
the solid angle, while the bars indicate the range spanned by
projection effects. Solid and dashed lines represent the $\Mbh$-$\sg0$
relation for $\alpha=4$ and $\alpha=5$, respectively, while vertical
dotted lines show the observed scatter around these best fits. {\it
Panel b}: same data as in panel a, but for maximally radiative BH
merging. {\it Panel c}: stellar mass vs. BH mass for classical BH
merging; $M_{*,0}$ is the stellar mass of the first progenitor and the
solid line represents the Magorrian ($\Mbh \propto M_{\rm bulge}$)
relation. {\it Panel d}: same data as panel c, but for maximally
radiative BH merging. Symbols are the same as in Fig. 2.}
\end{center}
\end{figure*}

On the basis of the results of the previous Section, it is
particularly interesting to investigate whether dissipationless
merging is able to reproduce the $\Mbh$-$\sg0$ relation, and we
attempt to answer this question by using the simulations presented in
Section 3. Here we recall that in our simulations we do not take into
account the presence BH in the merging galaxies, yet we argue that we
can reach robust conclusions, and this for two reasons. In fact, as
well known, at the equilibrium (i.e. after the virialization of the
end--product) the presence of the BH has not significant influence on
$\sg0$. This can easily be seen by considering that the sphere of
influence of a BH with mass $\Mbh$ at the center of a galaxy with
central velocity dispersion $\sg0$ has a fiducial radius $\rbh \equiv
G \Mbh / \sg0^2$, and combining this equation with equation (7) we
obtain
\begin{equation}
\rbh \simeq 0.01 \times \left({\sg0 \over 200\,{\rm
km\,s^{-1}}}\right)^{\alpha-2}\,{\rm kpc},  
\end{equation}
where we assumed $\Mbh = 10^8M_{\odot}$ when $\sg0 = 200\,{\rm
km\,s^{-1}}$ (see, e.g., Tremaine et al. 2002).  The central velocity
dispersion used in the definition of the FP and of the $\Mbh$-$\sg0$
relations is the luminosity weighted projected velocity dispersion
inside the radius $\cRe/8$ which, for ellipticals with $\sg0 \simeq
200$ km s$^{-1}$, is in the range $0.2 \, \lsim\, \cRe/8 \, \lsim \,
0.6$ kpc, one order of magnitude larger than $\rbh$.  Thus, once the
numerical end--products reached the equilibrium, their $\sg0$ is a
good estimate of the real quantity even in presence of a BH of
realistic mass. The second reason requires a more careful
discussion. In fact, during the dynamical evolution of the merging
system, a BH binary is formed and, in principle, its evolution could
affect the kinematic properties of the end--product even beyond
$r_{\rm BH}$.  However Milosavljevic \& Merritt (2001), with the aid
of numerical simulations, showed that the formation of a BH binary
does not modify significantly the central velocity dispersion measured
within the standard aperture $\cRe/8$, although the inner density
profile {\it is} modified (for interesting observational evidences of
this case see Lauer et al. 2002). In any case, we remark here that, if
any, the expected effect of binary BHs is to {\it decrease} the
central velocity dispersion, as a consequence of dynamical friction
heating against background stars.

On the basis of these considerations we simply assume that each seed
galaxy contains a BH of mass $\Mbhzero$, and that each merging
end--product contains a BH obtained by the merging of the BHs of the
progenitors (at the end of this Section, however, we will discuss two
important problem faced by this last assumption). Unfortunately, BH
merging is still a poorly understood physical process, in particular
with respect to the amount of emitted gravitational waves (see, e.g.,
CvA, and references therein), and for this reason we consider two
extreme situations: the case of {\it classical} combination of masses
($M_{\rm BH,1+2}=M_{\rm BH,1}+M_{\rm BH,2}$, with no emission as
gravitational waves), and the case of {\it maximally efficient
radiative merging} ($M^2_{\rm BH,1+2}=M^2_{\rm BH,1}+M^2_{\rm BH,2}$,
corresponding to entropy conservation in a merging of two non rotating
BHs). Following this choice, in Fig. 6 we plot the central velocity
dispersion of the mergers versus the mass of their central BH in the
case of classical (panel a) and maximally radiative (panel b) BH
merging. As expected from the similarity between the FJ and the
$\Mbh$-$\sg0$ relations, in the classical case, both (one--component)
equal mass mergers (solid squares and triangles) and accretion mergers
(empty circles and triangles) are unable to reproduce the observed
relation, even when its slope is assumed as high as 5, with
representative points well outside the observed scatter (vertical
dotted bars). As it happens for the FJ relation, the situation is
somewhat better when considering equal mass mergers of two--component
galaxy models: in presence of DM the stellar central velocity
dispersion is higher for given BH mass and the end--products are found
closer to the observed relation (stars in Fig. 6). Thus, as for the
FJ, the reason of the failure of dissipationless merging at
reproducing the $\Mbh$-$\sg0$ relation is that the end--products are
characterized by a too low $\sg0$ for the given $\Mbh$, i.e., $\Mbh$
is too high for the resulting $\sg0$. A promising solution to this
problem could be the emission of some fraction of $\Mbh$ as
gravitational waves. In fact, by assuming maximally efficient
radiative BH merging, the points are found remarkably closer to the
observed relation, even if it is apparent that the slope of the
$\Mbh$-$\sg0$ relation is not well reproduced by the end--products of
head--on accretion (empty circles). Note however that, while in the
classical scenario the Magorrian relation is (obviously) nicely
reproduced (Fig. 6), in case of substantial emission of gravitational
waves the relation between BH mass and bulge mass is {\it not}
reproduced (Fig. 6).

All the results presented in this Section holds under the strong
assumption that the BHs of the merging galaxies are retained by the
end--products, but there at least two basic mechanisms that could be
effective in expelling the central BHs.  The first is related to the
general instability of three body systems: if a third galaxy is
accreted by the end--product of a previous merging before the binary
BH at its center merged in a single BH, then the escape of the
smallest BH is highly possible. It is clear that if this process
happens more than a few times, then the Magorrian relation will be not
preserved at the end. A plausible solution to this problem is then to
assume that the characteristic time of BH merging (first by dynamical
friction on the background stars and then by emission of gravitational
waves) is shorter than the characteristic time between two galaxy
merging (for a detailed discussion of this problem see, e.g.,
Milosavljevic \& Merritt 2001, Yu 2002, Haehnelt \& Kauffmann 2002,
Volonteri, Haardt \& Madau 2003). However, a second physical mechanism
could be more effective in expelling the resulting BH from the center
of a galaxy merger, i.e., anisotropic emission of gravitational
waves. This process, commonly known as the ``kick velocity'', is
directly related to the fraction of BH mass emitted anisotropically
during the BH coalescence (see, e.g., Flanagan \& Hughes 1998,
CvA). Gravitational waves in fact travel at the velocity of light, and
so even the anisotropic emission of {\it a few thousandths} of the
mass of the BH binary will produce a recoil (due to linear momentum
conservation) of the resulting BH with a characteristic velocity
higher than the escape velocity typical of massive galaxies. In
conclusion, it seems to us not trivial to assume that in each galaxy
merging the resulting BH will reside at the center (see, e.g.,
Haehnelt \& Kauffmann 2000).

\section{Discussion and conclusions}

As outlined in the Introduction, the present study is similar, in some
aspects, to previous studies concerning the effects of merging on the
FP. The results of our simulations are consistent with those of
Capelato et al. (1995), based on N-body simulations of King (1966)
galaxy models, and of Dantas et al. (2003), based on Hernquist models,
who found that the edge--on FP is reproduced by dissipationless
merging, owing to non--homology of the end--products.  However, they
considered only the first two or three steps of the merging
hierarchy. In the same line of investigation, Gonzalez-Garcia \& van
Albada (2003) found that also the first generation of end--products of
dissipationless merging based on Jaffe (1983) models preserves the FP
when seen edge--on. At variance with these works, we used galaxy
models with a substantially larger number of particles, we explored
the cumulative effects of several generations of merging, and we also
compared with other scaling laws, such as the Faber-Jackson, the
Kormendy, and the $\Mbh$-$\sg0$ relations.

The main results of our simulations can be summarized as follows:
\begin{itemize}

\item The edge--on FP is well reproduced by dissipationless
hierarchical equal mass merging of one and two--component galaxy
models, and by accretion simulations with substantial angular momentum,
with their seed galaxies (i.e. the merging zeroth order generation)
placed on the FP itself. On the contrary, in the case of head--on
accretion, the $\kt$ coordinate {\it decreases} at increasing $\ku$,
at variance with real galaxies. The physical reason of these different
behavior is due to a different evolution of structural non--homology
of the end--products in the two scenarios; in fact the Sersic best fit
parameter $m$ grows monotonically with mass in equal mass mergers (and
accretion with angular momentum), while {\it decreases} with mass in
head--on accretion end--products.

\item In both scenarios the scatter in the ($\ku,\kt$) coordinates
associated to projection effects is of the order of the observed
1-$\sigma$ dispersion of the edge--on FP.

\item At variance with the results of the edge--on FP, all the results
of our simulations are in qualitative agreement when plotted in the
($\ku,\kd$) plane, where the FP is seen nearly face--on. In this plane
the mergers move parallel to the line defining the zone of avoidance,
and remain in the region of the FP populated by real galaxies.

\item As a direct consequence of the numerical findings that the
``observables'' $\sg0$ and $\cRe$ follow closely the evolution of
$\sgv$ and $\rv$ as predicted by the virial theorem, the end--products
of our simulations fail to reproduce both the FJ and Kormendy
relations. However, in the case of equal mass merging and accretion
merging with angular momentum, the large values of $\cRe$ and the
nearly constant values of $\sg0$ curiously compensate, and the
end--products follow nicely the edge--on FP (see the first point
above).

\item Under the reasonable hypothesis that the derived values of
$\sg0$ are not strongly affected by the dynamical evolution of binary
BHs, our results show that dissipationless merging, while in
accordance with the Magorrian relation, fails to reproduce the
$\Mbh$-$\sg0$ relation. We have then shown that, allowing for
substantial emission of gravitational waves during the BHs
coalescence, the $\Mbh$-$\sg0$ relation is surprisingly reproduced,
but the Magorrian relation is not. Merging by head--on accretion
suffers of an additional problem when considered together with the
Magorrian relation: our simulations in fact showed that the best
Sersic $m$ parameter decreases at increasing mass of the end--product,
a behavior opposite to what is empirically found (Graham et al. 2001).

\end{itemize}

In conclusion the results of the presented simulations suggest that
{\it substantial} dissipationless merging (especially when involving
stellar systems of very different mass) cannot be the basic mechanism
of formation of Es. We note, however, that our exploration of the
parameter space is by no means complete, because in the simulations we
did not consider initial conditions corresponding to {\it multiple}
merging. In any case, indications exist that our results should be
obtained also under these more general circumstances (Nipoti et
al. 2003b). Finally, we note that from the numerical results one cannot
exclude, on the basis of FP, FJ, and Kormendy relations, that Es could
experience {\it few} occasional mergings with other Es, even in recent
times as observed by van Dokkum et al. (1999). In any case, it is well
known that many other astrophysical evidences, based on stellar
population properties, such as the ${\rm Mg}_2$-$\sg0$ and the
color-magnitude relations, and the waveband dependence of the relation
between mass--to--light ratio and galaxy luminosity (see, e.g., Pahre
et al. 1998b, Pahre, de Carvalho \& Djorgovski 1998a), strongly argue for
a substantial dissipative phase in the formation of spheroids.

If we exclude dissipationless merging as the formation mechanism for
Es, then we are left with two possible solutions, namely the
monolithic scenario or a merging scenario where dissipation plays an
important role. In this last case the presented simulations can also
say something. In fact gas dissipation should be effective in
shrinking the density distribution of the merging end--products, and,
at the same time, increasing their central velocity dispersion,
changes that are in the direction required by the FJ and Kormendy
relations.  In addition, in order to preserve the Magorrian relation,
a fraction of the star forming dissipating gas must flow on the
central BH, producing QSO activity; in this case the cosmological
evolution of QSOs would be a tracer of the merging history, in
accordance with the old ages of stars in Es. As a consequence,
numerical simulations of galaxy merging in presence of gas (see, e.g.,
Bekki 1998), taking also into account the feedback from the central BH
in a self-consistent way (see, e.g., Tabor \& Binney 1993, Binney \&
Tabor 1995, Ciotti \& Ostriker 1997, 2001), and finally a better
understanding of BH merging (see, e.g. Merritt \& Ekers 2002; Hughes
\& Blandford 2002) are highly needed for a substantial progress in
this field.

\section*{Acknowledgments}

L.C. would like to thank Giuseppe Bertin, Piero Madau, Jerry Ostriker,
Silvia Pellegrini, Alvio Renzini and Tjeerd van Albada for useful
discussions. We thank Reinaldo de Carvalho and the anonymous Referee
for helpful comments. C.N. and P.L. are grateful to CINECA (Bologna)
for assistance with the use of the Cray T3E and of the IBM Linux
Cluster. This work was supported by MURST CoFin2000.

\end{document}